\documentclass{kapproc} 

\usepackage{procps} 

\usepackage[dvips]{graphicx}

\upperandlowercase

\begin{document}

\articletitle{Efficiency of the dynamical mechanism}

\articlesubtitle{}

\author{F. Combes}
\affil{LERMA, Observatoire de Paris\\
61 Av. de l'Observatoire, F-75014 Paris, France}
\email{francoise.combes@obspm.fr}

\begin{abstract}
The most extreme starbursts occur in galaxy mergers,
and it is now acknowledged that dynamical triggering
has a primary importance in star formation.
This triggering is due partly to the enhanced velocity
dispersion provided by gravitational instabilities,
such as density waves and bars,
but mainly to the radial gas flows they drive, allowing
large amounts of gas to condense towards nuclear
regions in a small time scale.
Numerical simulations with several gas phases,
taking into account the feedback to regulate
star formation, have explored the various processes, 
using recipes like the Schmidt law, moderated by
the gas instability criterion.  May be the most fundamental
parameter in starbursts is the availability of gas:
this sheds light on the amount of external gas accretion
in galaxy evolution. The detailed
mechanisms governing gas infall in the inner parts of 
galaxy disks are discussed.
\end{abstract}


\section{Introduction}

The most spectacular evidence for dynamical triggering
of starbursts is that
ULIRGs are all mergers of galaxies (e.g. Sanders \& Mirabel 1996).
They have much more gas, dust and young stars then normal 
spiral galaxies, but they are quite rare objects in the 
nearby universe. 
On the contrary, interacting galaxies do not show intense starbursts
(e.g. Bergvall et al 2003, but see Barton et al 2000, Nikolic et 
al. 2004), or only in their centers. From
many observational studies, it appears that galaxy 
interactions are a necessary condition, but not a sufficient
condition to trigger a starburst.
Another necessary condition of course is the presence of 
large amounts of gas.

For small systems, interactions are even not necessary,
since spontaneous star-formation can occur intermittently in
bursts. Starbursting dwarf galaxies have no excess of companions
(Telles \& Maddox 2000, Brosch et al 2004, except Blue Compact 
Dwarfs according to Hunter \& Elmegreen 2004).
So for dwarf galaxies, tides are not very important.
It is possible instead that the gas in these objects, 
having been accreted recently, is not yet in
dynamical equilibrium: observed asymmetries
could be due to sloshing gas inside dark haloes. 

\begin{figure}[ht]
\begin{center}
  \includegraphics[width=.6\textwidth]{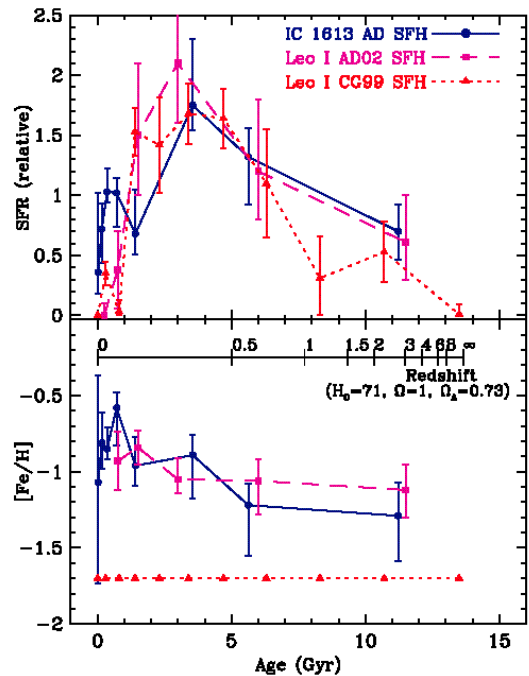}
\end{center}
  \caption{Star formation and metal
enrichment histories derived for IC1613 and Leo I dwarfs
by Skillman et al (2003). It is remarkable
that the bulk of the star formation
and metal enrichment has occurred since z=1. } 
\label{skil03}
\end{figure}

It is possible today to trace the star formation and 
chemical enrichment history of nearby galaxies, by studying
their stellar populations in detail and their metallicity.
The star formation history in the Small Magellanic Cloud
reveals some bursts corresponding to pericenters
with the Milky Way (Zaritsky \& Harris 2004).
The tidal-induced fraction of star formation could be
between 10 and 70\%. A good fit is impossible however
without large amounts of gas infall, at least 50\%.
 For two local dwarfs, Skillman et al (2003) conclude
also that the bulk of star formation is recent,
unlike the predictions of an exponentially
decreasing star formation history, if the system
had acquired most of its mass in early times
(Fig. \ref{skil03}).

\section{Dynamical Processes}

Empirically, star formation is observed to obey a 
global Schmidt law, where the rate of SF per unit surface
is a power n=1.5 of the average gas surface density in a
galaxy (e.g. Kennicutt 1998). It is remarkable that
this law holds with the same slope and is continuous,
for interacting and non-interacting objects, pointing
to the gas supply as the main factor.

\begin{figure}[ht]
  \includegraphics[width=.9\textwidth]{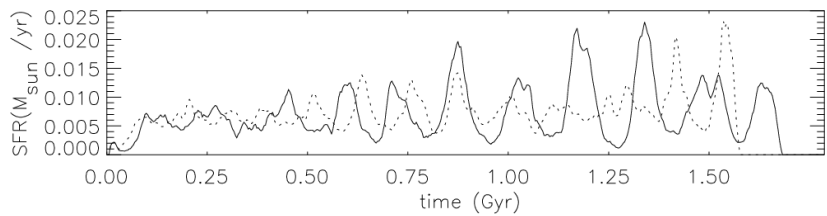}
  \caption{Star Formation History of a simulated dwarf galaxy.  
The dotted line indicates the SFR for a run with  50\% reduced feedback strength
(from Pelupessy et al 2004).}
\label{pelup04}
\end{figure}

This empirical law can be interpreted through several
processes: Jeans instability, since the SFR is
then proportional to the density $\rho$ and inversely
proportional to the dynamical time  
in $\rho^{-1/2}$, or cloud-cloud collisions (Elmegreen 1998),
contagious star formation, associated with feedback 
(generating chaotic conditions), etc.
These processes are able, without dynamical trigger,
to yield episodic bursts of star formation, and this is
well suited to dwarf galaxies, see Fig. \ref{pelup04}
(K\"oppen et al 1995, Pelupessy et al 2004)

For larger systems, large-scale dynamical instabilities must be invoked.
Density waves, in creating shocks and concentrations of mass in 
spiral arms, can favor star formation, but starbursts require
to gather large amounts of gas in a small area.
Radial gas flows due to bars, or spirals torques are then at work,
leading to molecular gas concentrations, and  circumnuclear starbursts
(e.g. Buta \& Combes 1996, Sakamoto et al 1999, Knapen 2004).
In galaxy clusters, star formation could be induced by shocks
with the intra-cluster medium (Bekki \& Couch 2003).

\subsection{ULIRGS}

Ultra-Luminous InfraRed Galaxies
 have not only more gas and star formation, but also an enhanced
 star formation efficiency (SFE) defined as the ratio
of SFR traced by the far infrared luminosity to the 
available fuel, traced by the CO emission (for the H$_2$ gas). 
More generally, in interacting galaxies, the 
CO emission relative to blue luminosity is multiplied by 5
and more concentrated (Braine \& Combes 1993). This 
certainly means that the H$_2$ content is larger;  the
interpretation in terms of a lower CO-to-H$_2$ conversion factor   
would lead to an excessive star formation efficiency  
SFE=L(FIR)/M(H$_2$).  

These enhanced gas amount and concentration
 can be explained by the gravitational torques of the interactions
driving gas very quickly to the centers.
Gas in ULIRGs is concentrated in central nuclear disks or 
rings (Downes \& Solomon 1998).
The condition to have a starburst is to accumulate gas in a time short enough
that feedback mechanisms have no time to regulate.
Also, the tidal forces are generally compressive in the centers,
which favors cloud collapse.

\subsection{Compressive tidal forces}

For a spherical density profile,
modelled as a power-law  $\rho (r) \sim r^{-\alpha}$, the  corresponding
acceleration is in $r^{1-\alpha}$, so the gravitational
attraction can increase with distance from center,
if  $0 <  \alpha  <  1 $. Therefore the
 tidal force is then compressive: $Ftid \sim (1-\alpha) r^{-\alpha}$,
in particular, for a core with constant 
density ($\alpha=0$). The rotation curve 
$V_{rot}$ in $r^{1-\alpha/2}$, would then be almost rigid
rotation.
Molecular clouds inside the core are then compressed, and star
formation can be triggered.

This phenomenon can also explain the formation of nuclear 
starbursts and then young nuclear stellar disks in
some barred galaxies.
Decoupled stellar nuclear disks are frequently observed
in double-barred Seyfert galaxies (Emsellem et al 2001).
The observed velocity dispersion reveals a
characteristic drop in the center. The
proposed interpretation invokes star formation
in a decoupled nuclear gas disk 
(Wozniak  et al 2003).

\subsection{Star formation recipes}

\begin{figure}[ht]
   \centering
\begin{minipage}{4cm}
\rotatebox{-90}{\includegraphics[width=7cm]{combesf-fig3a.ps}}
\end{minipage}
\begin{minipage}{4cm}
{\includegraphics[height=7.5cm]{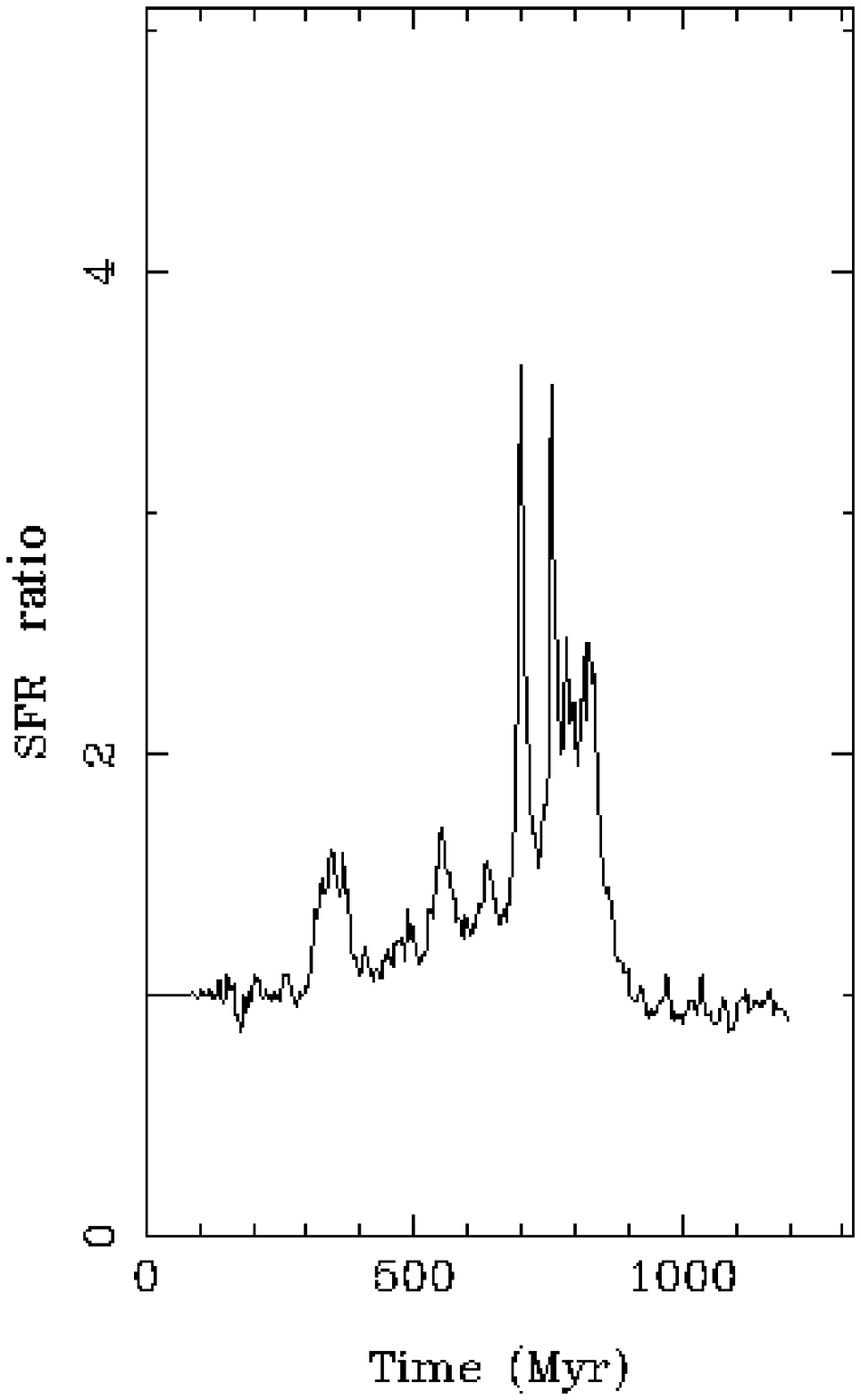}}
\end{minipage}
      \caption{ Star formation history during a major
merger of two Sb spiral galaxies:
{\bf Left}, the star formation rate versus time, showing
the global exponential decline; 
{\bf Right}, the ratio between SFR in the merger run and
the corresponding control run with the two galaxies isolated.
}
       \label{galmer}
   \end{figure}

Numerical simulations use recipes for star
formation and feedback phenomena, since this is
 sub-grid physics (Katz 1992, Mihos \& Hernquist 1994, 96).
These recipes include the
Schmidt law with exponent n=1.5, together
with a gas density threshold.
The star formation rate is however generally
decreasing exponentially with time, in isolated
system, even taking into account stellar mass loss
(see Fig. \ref{galmer}). 
When comparing the star formation history in an
isolated galaxy with respect to a merger, the
exponential law dominates, unless the SFR is 
normalised to the isolated case.

According to the detailed geometry, mass ratios or
dynamical state of the merging galaxies,
star formation can be delayed until the final merger,
but the availability of gas is the main issue.

\subsection{Importance of gas accretion}

Galaxies in the middle of the Hubble sequence have experienced about
constant SFR along their lives 
(Kennicutt et al 1994, SDSS: Brinchmann et al 2004).
The study of stellar populations in the large SDSS sample 
has shown that only 
massive galaxies have formed most of their stars at early times,
while dwarfs are still forming now
(Heavens et al 2004). 
Only intermediate masses have in average maintained
their star formation rate over a Hubble time.

Even taking into account the stellar mass loss, an isolated 
galaxy should have an exponentially decreasing  star
formation history.
Galaxies must therefore accrete large amounts of gas mass
along their lives, to fuel star formation.

Large amounts of gas accretion
is also required to explain the observed bar frequency
today (Block et al 2002). Numerical simulations
reveal that bars in gaseous spirals are quickly
destroyed, and only gas accretion can trigger
their reformation (Bournaud \& Combes 2002). 
To have the right frequency of bars at the present time, 
gas accretion must double the galaxy mass in 10Gyr.
This gas cannot come from
dwarf companions: they can provide at most 
10\% of the required gas and their dynamical interactions 
heat the disk. 
What is required is continuous cold gas accretion,
which could come
from the cosmological filaments in the near 
environment of galaxies.

\section{Conclusion}

Star formation depends essentially on the gas supply.
External gas accretion is essential for the efficiency 
of dynamical triggering.
Galaxy interactions, and the accompanying bars
and spirals, help to drive the accreted gas radially inwards
 and trigger central starbursts.

Gas accretion regulates not only the star formation
history in galaxies, 
but also their dynamics (bars, spirals, warps, m=1...),
since bars require gas to reform, in a self-regulating
process.
 
According to the environment, hierarchical merging or 
secular evolution prevail.
In the field, accretion is dominant, and explains bars and spirals, and
the constant star formation rate for intermediate types.
In rich environments, galaxy evolution is faster, interactions
and mergers are much more important;
secular evolution of galaxies is halted at $z \sim 1$, since galaxies
are stripped from their gas reservoirs.


\end{document}